\documentclass[12pt,draft]{article}
\usepackage{amssymb,euscript}

\newcommand{\ie}{{\it i.e.}}
\newcommand{\eg}{{\it e.g.}}
\newcommand{\e}{\begin{equation}}
\newcommand{\ee}{\end{equation}}
\newcommand{\ea}{\begin{eqnarray}}
\newcommand{\eea}{\end{eqnarray}}
\newcommand{\nn}{\nonumber}
\newcommand{\p}{\partial}

\newcommand{\Tr}{{\rm Tr}}

\begin{document}

\begin{flushright}
\end{flushright}
\begin{flushright}
\end{flushright}
\begin{center}

{\LARGE {\sc Comments On Multiple M2-branes\\}}

\bigskip
\bigskip

{\sc Jonathan Bagger\footnote{bagger@jhu.edu}} \\
{Department of Physics and Astronomy\\
Johns Hopkins University\\
3400 North Charles Street\\
Baltimore, MD 21218, USA }
\\
\bigskip
and
\\
\bigskip
{\sc Neil Lambert\footnote{neil.lambert@kcl.ac.uk}} \\
{Department of Mathematics\\
King's College London\\
The Strand\\
London WC2R 2LS, UK\\}

\end{center}

\bigskip
\begin{center}
{\bf {\sc Abstract}}
\end{center}
Recently a three-dimensional field theory was derived that is
consistent with all the symmetries expected of the worldvolume
action for multiple M2-branes. In this note we examine several
physical predictions of this model and show that they are in
agreement with expected M2-brane dynamics. In particular, we discuss
the quantization of the Chern-Simons coefficient, the vacuum moduli
space, a massive deformation leading to fuzzy three-sphere vacua, and
a possible large $n$ limit. In this large $n$ limit,
the fuzzy funnel solution correctly reproduces the mass of an
M5-brane.

\newpage

\section{\sl Introduction}

M-branes are mysterious objects (see \eg\ \cite{Berman:2007bv}) and
virtually nothing is known about their underlying dynamics beyond
the case of a single brane. This is in sharp contrast to D-branes
\cite{Polchinski:1995mt}, where a microscopic description in terms
of open strings has driven a huge amount of progress in string
theory and gauge theory.

In \cite{Bagger:2006sk} a model for multiple M2-branes was proposed
in which the scalar fields take values in an algebra that admits a
totally antisymmetric tri-linear product.  It was conjectured that
this model could be made maximally supersymmetric by including a
non-propagating gauge field.  The corresponding supersymmetry
algebra was shown to close in \cite{Gustavsson:2007vu}(v4) and
then in \cite{Bagger:2007jr}, where the full equations of motion and
Lagrangian were given.  The theory is consistent with all the
symmetries expected from multiple M2-branes.

Given a theory with the symmetries of a multiple M2-branes, it is
natural to see if it reproduces other properties expected of
such systems. In this paper we will consider various
consequences of the Lagrangian presented in \cite{Bagger:2007jr}. We
will find that several predictions are consistent with expectations,
although some aspects of the algebra are not sufficiently well
developed to check them all.

The rest of this paper is organized as follows. In section 2 we
review the results of Ref.~\cite{Bagger:2007jr}.  In section 3
we show that consistency requires quantization of the structure
constants associated with the tri-linear product.  This suggests that
the theory is conformally invariant to all orders in perturbation
theory.  In section 4 we examine the vacuum moduli space of the
simplest nontrivial model and argue that, surprisingly, it describes
three M2-branes.  In section 5 we consider a mass deformation of
the M2-brane worldvolume and show that it leads to fuzzy sphere
vacua, as argued in \cite{Bena:2000zb}.  In section 6 we propose
an algebra for an infinite number of M2-branes and show that, when
combined with the quantization conditions in section 3, it
reproduces the correct energy for the supersymmetric fuzzy funnel
solutions of \cite{Basu:2004ed}.  In the appendix we demonstrate
that the approaches of \cite{Gustavsson:2007vu} and
\cite{Bagger:2007jr} are equivalent, despite their apparently
different algebraic structures.

\section{\sl The Field Theory}

The field theory derived in \cite{Bagger:2007jr} assumes
that the scalars $X^I$, $I=3,4,...,10,$ and fermions $\Psi$,
$\Gamma_{012}\Psi=-\Psi$, take values in a so-called
three-algebra $\cal A$.  This is a vector space with
basis $T^a$, $a=1,...,N$, that is endowed with a trilinear
antisymmetric product
\begin{equation}
[T^a,T^b,T^c] = f^{abc}{}_{d}\,T^d,
\end{equation}
from which is it is clear that $f^{abc}{}_{d}=f^{[abc]}{}_{d}$.
We further suppose there is trace-form that provides a metric
\begin{equation}
h^{ab} = \Tr(T^a,T^b),
\end{equation}
which we assume to be positive definite. This allows us to raise and lower indices: $f^{abcd} =f^{abc}{}_{e}h^{ed}$.

We require two conditions on the triple product.  The first is the fundamental identity
\begin{eqnarray}
[T^a,T^b,[T^c,T^d,T^e]] &=&[[T^a,T^b,T^c],T^d,T^e]
+[T^c,[T^a,T^b,T^d],T^e]\\
\nn && +\ [T^c,T^d,[T^a,T^b,T^e]],
\end{eqnarray}
for all $a,b=1,..,N$. This is equivalent to
\begin{equation}\label{fID}
f^{efg}{}_{d}f^{abc}{}_{g}=f^{efa}{}_{g}f^{bcg}{}_{d}
+f^{efb}{}_{g}f^{cag}{}_{d}+f^{efc}{}_{g}f^{abg}{}_{d}.
\end{equation}
The second is
\begin{equation}\label{trace}
\Tr(T^a,[T^b,T^c,T^d]) = -\Tr([T^a,T^b,T^c],T^d),
\end{equation}
for all $a,b=1,..,N$.  This implies that the $f^{abcd}$ are totally antisymmetric,
\begin{equation}\label{antisym}
f^{abcd}= f^{[abcd]}.
\end{equation}

We augment this algebra by including an element $T^0$ that has a vanishing triple product with everything, \ie\ that satisfies $f^{0ab}{}_{d}=0$.  Assuming $h^{0b}=0$ when $b\ne 0$, we find $f^{abc}{}_{0}=0$.  Thus this mode decouples and can be
interpreted as the centre-of-mass coordinate.

There is a natural gauge symmetry on the fields $X^I_d$, where $\delta X^I_d = \Lambda_{ab}f^{abc}{}_{d}X^I_c\equiv \tilde\Lambda^c{}_{d}X^I_c$. There is a covariant derivative $D_\mu X^I_d =\partial_\mu X^I_d - \tilde A_\mu{}^{c}{}_{d}X^I_c$, with $\delta \tilde A_\mu{}^{c}{}_{d} = D_\mu \tilde \Lambda^c{}_{d} $, as well as a gauge-covariant field strength $\tilde F_{\mu\nu}{}^c{}_d$.  The space of all $\tilde \Lambda^c{}_d$ is closed under the ordinary matrix commutator, so it generates a matrix Lie algebra $\cal G$. {}From this perspective, $\tilde A_{\mu}{}^c{}_d$ is the usual gauge connection in the adjoint representation of $\cal G$, while the elements of $\cal A$ are in the fundamental representation. The fundamental identity implies that $f^{abcd}$ is an invariant 4-form of $\cal G$.

The Lagrangian derived in \cite{Bagger:2007jr} is
\begin{eqnarray}\label{action}
\nonumber {\cal L} &=& -\frac{1}{2}D^\mu X^{aI} D_\mu X^{I}_{a}
+\frac{i}{2}\bar\Psi^a\Gamma^\mu D_\mu \Psi_a
+\frac{i}{4}\bar\Psi_b\Gamma_{IJ}X^I_cX^J_d\Psi_a f^{abcd}\\
&& - V(X)+\frac{1}{2}\varepsilon^{\mu\nu\lambda}\left(f^{abcd}A_{\mu
ab}\partial_\nu A_{\lambda cd} +\frac{2}{3}f^{cda}{}_gf^{efgb}
A_{\mu ab}A_{\nu cd}A_{\lambda ef}\right),\nn\\
\end{eqnarray}
where
\begin{eqnarray}\label{potential}
V (X) &=& \frac{1}{12}{\rm Tr}([X^I,X^J,X^K],[X^I,X^J,X^K])
\end{eqnarray}
and $\tilde A_{\mu}{}^c{}_d = A_{\mu ab}f^{abc}{}_{d}$.  The Lagrangian is invariant, up to boundary terms, under the transformations
\begin{eqnarray}\label{susygauged}
\nonumber \delta X^I_a &=& i\bar\epsilon\Gamma^I\Psi_a\\
\delta \Psi_a &=& D_\mu X^I_a\Gamma^\mu \Gamma^I\epsilon
-\frac{1}{6}
X^I_bX^J_cX^K_d f^{bcd}{}_{a}\Gamma^{IJK}\epsilon \\
\nonumber \delta\tilde A_{\mu}{}^b{}_a &=& i\bar\epsilon
\Gamma_\mu\Gamma_IX^I_c\Psi_d f^{cdb}{}_{a},
\end{eqnarray}
where $\Gamma_{012}\epsilon=\epsilon$. These transformations close
into translations and gauge transformations.  Thus the action has 16
supersymmetries. It also has a manifest $SO(8)$ R-symmetry that acts
on the scalars $X^I$. Furthermore, the action preserves parity if
$f^{abcd}$ is taken to be parity odd.  These are precisely the
symmetries that are expected of the worldvolume description of
multiple M2-branes.

This action provides a non-Abelian
generalization of the single M2-brane action and describes M2-branes
propagating in a flat eleven-dimensional spacetime.  As such, it
presumably arises as the lowest-order term in a derivative
expansion in static gauge of some new $\kappa$-symmetric action
that generalizes the Born-Infeld action of D-branes.  It would be
interesting to study this in more detail; however to date, non-Abelian
$\kappa$ symmetry is poorly understood.  Here we are compelled to
test the predictions of this model against various expectations for
multi M2-branes.

\section{\sl Quantizing $f^{abcd}$}

Classically, given any choice of structure constants that satisfies
the conditions of a three-algebra, namely (\ref{fID}) and
(\ref{antisym}), one can rescale the $f^{abcd}$ and preserve the
defining conditions.  In a quantum theory, however, the coefficient
of a Chern-Simons term must be quantized.  Therefore, for the case
at hand, we expect such a constraint on the $f^{abcd}$.

To proceed further, we observe that $f^{abcd}$ defines a linear map
acting on the vector space of antisymmetric $N\times N$ matrices,
\begin{equation}
f(M_{ab})  = \frac{1}{2}f_{ab}^{\ \ cd}M_{cd},
\end{equation}
where we use $h^{ab}$ and its inverse to raise and lower indices.
Using the natural inner product, $\langle M^1,M^2\rangle =
M^1_{ab}M^{2ab}$, one sees that the map is real and symmetric.
Therefore it can be diagonalized with eigenvalues that we denote by
$\lambda$. Using the fundamental identity, one can show that if
$M^1_{ab}$ and $M^2_{ab}$ are matrices with eigenvalues $\lambda_1$
and $\lambda_2$, respectively, then
\begin{equation}\label{eigenID}
f([\tilde M^1,\tilde M^2]) = \lambda_2[\tilde M^1,\tilde M^2],
\end{equation}
where $\tilde M^c{}_d = f^{abc}{}_{d}M_{ab}$ and $[\tilde M^1,\tilde M^2]$ is the ordinary matrix commutator. Thus each eigenspace of $f$ is closed under commutation and defines a Lie subalgebra ${\cal G}_\lambda \subset {\cal G}$.

It follows from (\ref{eigenID}) that
\begin{equation}
f([\tilde M^1,\tilde M^2]) = - f([\tilde M^2,\tilde M^1]) =
-\lambda_1 [\tilde M^2,\tilde M^1] =\lambda_1 [\tilde M^1,\tilde
M^2],
\end{equation}
and hence
\begin{equation} \lambda_1 [\tilde M^1,\tilde M^2] =
\lambda_2 [\tilde M^1,\tilde M^2] .
\end{equation}
This shows that  $[\tilde M^1,\tilde M^2]=0$ if $\lambda_1\ne\lambda_2$. Thus the various sub-algebras ${\cal G}_\lambda$ commute and $\cal G$ decomposes as
\begin{equation}
{\cal G} = \oplus_\lambda{\cal G}_\lambda .
\end{equation}

As a result of this fact, we can choose a basis in which the twisted Chern-Simons term is
\begin{equation}
\sum_{\lambda\ne 0} \frac{1}{4\lambda}{\rm Tr}\left(\tilde A^{(\lambda)}\wedge
d \tilde A^{(\lambda)} +\frac{2}{3}\tilde A^{(\lambda)}\wedge \tilde
A^{(\lambda)}\wedge \tilde A^{(\lambda)}\right) ,
\end{equation}
where $\tilde A^{(\lambda)}=\tilde A^{(\lambda)}_\mu dx^\mu$ is the projection of the gauge field onto the eigenspace ${\cal G}_\lambda$, and ordinary matrix multiplication is understood to apply.  It is well known that for the path integral to be well-defined, the coefficient of a Chern-Simons term must be $k/4\pi$, where $k\in {\mathbb Z}$ \cite{Deser:1982vy} is called the level.  Thus we see that the eigenvalues of $f$ must satisfy
\begin{equation}
\lambda = \frac{\pi}{k}
\end{equation}
for each $\lambda$, with $k\in {\mathbb Z}$.  In the quantum theory,
there is no freedom to rescale the $f^{abcd}$.  For simplicity, in
the rest of this paper we only consider the case $k=1$. It
would be interesting to examine the physical interpretations of
other values of $k$.

Note that the quantization of $f^{abcd}$ suggests that there are no
continuous parameters in the theory.  If so, the
theory must be conformally invariant to all orders in perturbation
theory; since there are no coupling constants, there are no
parameters to run. Supersymmetry determines them once and for all.

\section{\sl Vacuum Moduli Space}

To explore the connection between our theory and multiple
M2-branes, it is natural to start with the vacuum moduli space.
Setting $\tilde A_\mu=\Psi=\partial_\mu X^I=0$, the requirement
that all supersymmetries be preserved implies that
\begin{equation}\label{vaceq}
[X^I,X^J,X^K]=0 ,
\end{equation}
for all $X^I$. This condition also ensures that the equations of
motion are satisfied.

Let us focus on the simplest nontrivial possibility, in which the three-algebra $\cal A$ has four generators and hence, given the quantization condition found above,
\begin{equation}
f^{abcd} = {\pi}\varepsilon^{abcd} ,
\end{equation}
where $a,b=1,2,3,4$. Without loss of generality, we take $h^{ab} = \delta^{ab}$. We call this three-algebra ${\cal A}_4$; one can check that it satisfies the fundamental identity.  The solutions to the vacuum equations (\ref{vaceq}) are given by
\begin{equation}
X^I = a^I \alpha + b^I \beta ,
\end{equation}
where $\alpha$ and $\beta$ are any two elements of ${\cal A}_4$.

We next consider the gauge transformations. For the case at hand, the Lie algebra ${\cal G}_4$ is generated by $\varepsilon^{abc}{}_{d}\Lambda_{ab}$, where $\Lambda_{ab}$ is real and antisymmetric. Thus ${\cal G}_4$ is nothing but the set of all antisymmetric real $4\times 4$ matrices, \ie\ ${\cal G}_4=so(4)\equiv so(3)\oplus so(3)$. The elements $\alpha,\beta\in {\cal A}_4$ are in the fundamental representation of $SO(4)$. Therefore, up to a gauge transformation, we can set $\alpha \propto T^1$. Furthermore, using the little group $SO(3)$ of $T^1$ we can also choose $\beta\propto T^2$. Thus, up to a gauge transformation, the vacuum moduli space is parameterized by
\begin{equation}
X^I = a^IT^1 + b^I T^2 .
\end{equation}

This result implies that there are two bosonic zero modes for each
of the coordinates $X^I$. Including the overall center-of-mass
generator $T^0$, which decouples from all the interactions and gauge
symmetries, we find three bosonic zero modes for each scalar $X^I$.
As with multiple D-branes, M2-branes satisfy a no-force condition
and hence the most natural interpretation for these zero modes is
that they correspond to moving the M2-branes apart in transverse
directions. Therefore we are led to identify the Lagrangian with
the worldvolume theory of three M2-branes.  Note that this
argument assumes that our construction describes a generic point
in the moduli space; at
special points there may be fewer degrees of freedom.\footnote{We
are grateful to M.~van Raamsdonk for bringing this point
to our attention.}

It seems peculiar that the simplest nontrivial model describes three
M2-branes, rather than two. Let us therefore make some comments
as to why this might be the case. If we think of the worldvolume theory
of $n$ M2-branes as the IR fixed point of three-dimensional $U(n)$
super-Yang-Mills theory, then we expect fewer than $n^2$
degrees of freedom per field
in the IR. The smallest three-algebra must have at least four
generators, and including the center-of-mass gives a total of five
degrees of freedom per field. Thus one would not expect this algebra
to arise as the IR fixed point of two D2-branes, but rather as the fixed
point of three D2-branes. We note that the number $N$ of
degrees of freedom of $n$ M2-branes is conjectured to scale as $N =
n^{3/2}$ at large $n$, and it is encouraging to observe that
$3^{3/2}\sim 5.2$.

A possible resolution is that the IR fixed point
of two D2-branes is trivial.  It is natural
to expect that the worldvolume theory of two D2-branes loses at
least one degree of freedom in the IR, leaving at most three.
Factoring out the center-of-mass would then leave at most two interacting
degrees of freedom.  This might be too few to construct a theory that is
consistent with all the symmetries (even without assuming a
Lagrangian description). In particular, it is too small to identify
the fields with elements of a Lie algebra.

A more detailed analysis of the degrees of freedom requires
finding a class of three-algebras with arbitrarily large dimension
$N$. At present we do not know of any other finite-dimensional
cases.  However, we can make some observations. For $N>4$ the map
$f$ must have a nontrivial kernel, for the following reason. Suppose
it has a trivial kernel.  Then the space $\cal G$ of all $\tilde
\Lambda^c{}_{d} = f^{abc}{}_{d}\Lambda_{ab}$ would be all of
$so(N)$.  In that case $f^{abcd}$ would be an invariant four-tensor of
$so(N)$, but there are no such invariants for $N>4$. A non-vanishing
kernel would lead to additional vacuum moduli and hence a larger
number of M2-branes.

\section{\sl BPS States and a Mass Deformation}

In ref.~\cite{Bena:2000zb}, it was argued that in the presence
of a particular background four-form flux, M2-branes preserve
four supersymmetries and exhibit an $SO(4)$ R-symmetry.  Furthermore,
the flux induces a supersymmetric mass term for the worldvolume
scalars and fermions.  It was also argued that in this background,
the vacuum of $n$ M2-branes is a state in which the scalars
describe a fuzzy three-sphere in spacetime.  The M2-branes
`puff up' so that their worldvolume is of the form ${\mathbb R}^{1,2}
\times \tilde S^3$, where $\tilde S^3$ is a fuzzy three-sphere that
becomes a normal $S^3$ as $n\to\infty$.  This setup provides an
M-theory analog of the Myers effect that occurs for D-branes
in the presence of background fluxes \cite{Myers:1999ps}.

In this section we search for such solutions to our theory.
Since we are not interested in the gauge fields and fermions,
we truncate the Lagrangian to include only the scalar fields,
\begin{equation}
{\cal L}_B=-\frac{1}{2}\Tr\,(\partial_\mu X^I,\p^\mu X^I)
-\frac{1}{12}{\rm Tr}([X^I,X^J,X^K],[X^I,X^J,X^K]) .
\end{equation}
Consistency requires that $X^A_a\partial_\mu X^A_b f^{abc}{}_d=0$,
which follows from the gauge field equation of motion.  This relation is
satisfied in all the solutions discussed below.

We search for solutions with four non-vanishing scalars,
which we denote by $X^A$, $A=1,2,3,4$.  The search is
simplified by writing the potential in the following form,
\begin{equation}
V(X) =\frac{1}{2}\Tr\,(\p^A W,\p^{A} W),
\end{equation}
where
\begin{equation}\label{superP}
   {W} =
    \frac{1}{24}\varepsilon^{ABCD}{\rm Tr}(X^A,[X^B,X^C,X^D])
\end{equation}
is the `superpotential.'  We add an $SO(4)$ symmetric
mass term by generalizing (\ref{superP}) to
\begin{equation}
   {W} = \frac{1}{2}m{\rm Tr}(X^A,X^A) +
    \frac{1}{24 }\varepsilon^{ABCD}{\rm Tr}(X^A,[X^B,X^C,X^D]).
\end{equation}

Vacuum solutions require $\partial^A W=0$, or
\begin{equation}\label{vacua} m{X}^A
=-\frac{1}{6 }\varepsilon^{ABCD}[{ X}^B,{X}^C,{ X}^D] .
\end{equation}
In addition to the trivial solution $X^A=0$, eq.~(\ref{vacua}) has a
fuzzy $S^3$ solution in which the M2's puff up into a fuzzy three-sphere.
The two solutions describe two zero-energy vacuum states of the
M2-brane in the four-flux background.

To construct the fuzzy three-sphere vacuum, we suppose that the
three-algebra admits a representation of ${\cal A}_4$, so
the four generators $T^A$ satisfy $[{T}^A,{T}^B,{T}^C] =
\pi\varepsilon^{ABCD}T^D$.  The solution is found by
taking
\begin{equation}\label{FSvacua}
X^A = \sqrt{\frac{m}{\pi}}T^A ,
\end{equation}
with $m> 0$.  It describes a fuzzy three-sphere with radius
proportional to $\sqrt{m}$, in agreement with \cite{Bena:2000zb}.
In the case of D-branes, physically distinct vacua arise from
different representations of the symmetry algebra \cite{Myers:1999ps}.
Presumably, there is a similar family of solutions here, corresponding
to different numbers of M2-branes. We will not attempt to discuss
them further because we lack a sufficient understanding of
three-algebra representations.

We can also construct the BPS fuzzy funnel solutions of
\cite{Bagger:2006sk, Basu:2004ed}, in which the M2-branes
end on an M5-brane.  Following Bogomoln'yi,
we consider static solutions that depend on one coordinate
$x^2 = s$.  We write the energy as
\begin{eqnarray}
\nn E &=& \frac{1}{2}\int dsdx^1\Tr\left(\frac{dX^A}{ds} -
\partial^AW,\frac{dX^A}{ds} - \partial^AW\right) + 2
\partial^A W\frac{dX^A}{ds}\\
\nn &=&\frac{1}{2}\int dsdx^1\Tr\left(\frac{dX^A}{ds} -
\partial^AW,\frac{dX^A}{ds} - \partial^AW\right) +
2\frac{dW}{ds}.
\end{eqnarray}
Therefore, up to a boundary term, the minimum energy solutions satisfy
\begin{equation}\label{Bog}
\p_2  X^A =  \p^A W = mX^A +
    \frac{1}{6}\varepsilon^{ABCD}[{X}^B,{X}^C,{X}^D] .
\end{equation}
The fuzzy funnel solution is found by taking
\begin{equation}
X^A = f(s)T^A ,
\end{equation}
where $s=x^2$ and again the $T^A$ satisfy $[{T}^A,{T}^B,{T}^C] = \pi\varepsilon^{ABCD}T^D$. The equation for $f$ is
\begin{equation}
 f'=mf - \pi f^3 ;
\end{equation}
the solution is
\begin{equation}
 f =\sqrt{\frac{m}{\pi}}\frac{1}{\sqrt{1-ce^{ -2m s}}} ,
\end{equation}
where $c$ is a constant, which by translation can be set to $\pm 1$.

If $c=+1$ and $m>0$,  the solution behaves as $f =1/\sqrt{ 2\pi s}$
for small but positive $s$.  It approaches $f \to \sqrt{m/{\pi}}$ as
$s\to\infty$.  If $m<0$,  the function $f$ has the same behavior at
small and positive $s$, but $f\to 0$ as $s\to \infty$.  These
solutions describe fuzzy funnels in which an infinite radius fuzzy
three-sphere at $s=0$ relaxes into the fuzzy sphere or the trivial
vacuum, respectively, as $s\to \infty$.  The spacetime interpretation of
these solutions is that they correspond to M2-branes that end
on a single M5-brane, located at $s=0$ and infinitely
extended along the $(x^0,x^1,x^2,x^3,x^4,x^5)$ directions.

On the other hand, if $c=-1$ and $m>0$, the function $f$ is
bounded.  It vanishes exponentially as $s\to -\infty$ and
approaches $f\to \sqrt{m/{\pi}}$ as $s\to \infty$. Here there
is no divergent fuzzy funnel, {\it i.e.} no M5-brane.  This
solution smoothly interpolates between the trivial
and fuzzy sphere vacua. In other words, it is a traditional
domain wall that interpolates between two degenerate vacuum
solutions of the worldvolume effective action.

We conclude this section by explicitly checking the predictions
for the energy of the fuzzy funnel and the physical radius of the
fuzzy three-sphere vacuum.  We follow \cite{Basu:2004ed} and
calculate the total (divergent) energy of a fuzzy funnel solution,
with $m=0$,
\begin{eqnarray}
\nonumber E &=& \int_{-\infty}^\infty dx^1 \int^\infty_0 ds \Tr(\partial_2 X^A,\partial_2 X^A) \\
  &=& {\rm Tr}(T^A,T^A)\int_{-\infty}^\infty dx^1 \int^\infty_0 ds f'^2 \\
\nonumber  &=&\pi{{\rm Tr}(T^A,T^A)} \int_{-\infty}^\infty dx^1
\int^\infty_0  f^3 df .
\end{eqnarray}
Next we introduce the physical fuzzy sphere radius, $R$, which is defined to be the root mean square radius, averaged over the $n$ M2-branes,
\begin{eqnarray}
\nn R^2 &=&\frac{{{\rm Tr}(X^A,X^A)}}{nT_2}\\
&=& \frac{{{\rm Tr}(T^A,T^A)}}{nT_2}f^2 .
\end{eqnarray}
Note that we have inserted a factor of the membrane tension, $T_2$.  This follows from the fact that $X^A$ is canonically normalized and hence has mass dimension $1/2$.  Thus it cannot be directly interpreted as a spacetime coordinate. Instead, the spacetime coordinates should be identified with $X^A/\sqrt{T_2}$, which has the dimension of length.  This change of variable rescales the kinetic term of the action to
\begin{equation}\label{Rdef}
{\cal L} =- \frac{T_2}{2}\Tr( \partial_\mu X^I,\partial^\mu
X^I)+\ldots ,
\end{equation}
as expected for a membrane with tension $T_2$.

{}From these expressions we have what we need to compute the energy:
\begin{eqnarray}
\nonumber E &=& \frac{T_2^2}{2\pi} \frac{n^2}{{\rm Tr}(T^A,T^A)}
\int_{-\infty}^\infty dx^1\int_0^\infty 2\pi^2R^3dR  \\
  &=&T_5 \frac{n^2}{{\rm
Tr}(T^A,T^A)}\int d^5x ,
\end{eqnarray}
where we have used the fact that $T_2^2 = 2\pi T_5$ \cite{Schwarz:1995jq}. This expression, at least in the large $n$ limit, should reproduce the energy of an infinite M5-brane with tension $T_5$. This implies that
\begin{equation}\label{largenpredic}
\Tr(T^A,T^A) = n^2
\end{equation}
at large $n$. Unfortunately, we do not know enough about the
representations of three-algebras to confirm this prediction.

Finally, we return our attention to the fuzzy three-sphere vacuum described above.  Using (\ref{Rdef}), we see that in the large $n$ limit, the physical radius is
\begin{eqnarray}
R^2 &=& \frac{{{\rm Tr}(X^A,X^A)}}{nT_2} \nn \\
&=&\frac{m}{\pi}\frac{\Tr (T^A,T^A)}{nT_2} \nn \\[1mm]
&=& \frac{m n}{\pi T_2} ,
\end{eqnarray}
where we have used (\ref{largenpredic}).
In the units of \cite{Bena:2000zb}, the tension $T_2 = M_{11}^3/4\pi^2$, and hence
\begin{eqnarray}
R^2 = \frac{4\pi mn}{M_{11}^3}.
\end{eqnarray}
This agrees with the result in \cite{Bena:2000zb}, up to a factor $4/3$.

The energy density for the smooth domain wall that arises when $c=-1$ and $m>0$ can also be calculated.  We find
\begin{eqnarray}
\nn {\cal E} &=&   \int^\infty_{-\infty} ds \Tr(\partial_2 X^A,\partial_2 X^A)\\
&=& \Tr(T^A,T^A)\int^\infty_{-\infty} ds f'^2\\
\nn  &=& \frac{m^2}{4\pi}{\rm Tr}(T^A,T^A) \\
\nn &=& \frac{m^2n^2}{4\pi},
\end{eqnarray}
where the last line assumes the large-$n$ relation (\ref{largenpredic}).

\section{\sl The Large n Limit}

In this section we propose a large $n$ limit for the three-algebra
$\cal A$. A natural infinite-dimensional example of a three-algebra
is given by the space ${\cal C}^{\infty}(\Sigma)$ of differentiable
functions on a closed three-manifold $\Sigma$ endowed with a metric.
For simplicity we assume that $\Sigma$ is compact without boundary
and with a finite volume.  In this case the triple product is given by
the Nambu bracket \cite{Nambu:1973qe}
\begin{equation}\label{Nambu}
[X,Y,Z] = -\pi \star (dX\wedge dY\wedge dZ) .
\end{equation}
It can be shown that (\ref{Nambu}) satisfies the fundamental identity. Furthermore, if we take
\begin{equation}
\Tr (X,Y) =\frac{1}{{\rm vol}(\Sigma)} \int_\Sigma X\wedge \star Y ,
\end{equation}
then (\ref{Nambu}) also satisfies (\ref{trace}).  Note that we have normalized the trace-form so that the identity function has unit length (and can be identified with the translational generator $T^0$).

For this three-algebra, the gauge symmetry generated by the Nambu bracket is
\begin{eqnarray}
\nn\delta X &=& [\alpha,\beta,X]\\
 &=& v^k\partial_k X ,
\end{eqnarray}
where $v^k=-(\pi/\sqrt{g})\varepsilon^{ijk}\partial_i \alpha\partial_j\beta$ and  $\sigma^i$, $i=1,2,3$ are local coordinates on $\Sigma$. This transformation is nothing but an area-preserving diffeomorphism on $\Sigma$.

We wish to consider the large $n$ limit of the fuzzy three-sphere vacua found in the previous section. To do this we need to find a representation of ${\cal A}_4$ inside ${\cal C}^{\infty}(\Sigma)$. Since we need an $so(4)$ symmetry it is natural to take $\Sigma = S^3$, the unit sphere inside ${\mathbb R}^4$.  We then consider the four functions $T^A$ that describe the natural embedding of $S^3$ into ${\mathbb R}^4$:
\begin{eqnarray}
\nn  T^1 &=& \cos \theta_1 \\
\nn  T^2 &=& \sin\theta_1\cos\theta_2 \\
     T^3 &=& \sin\theta_1\sin\theta_2\cos\theta_3 \\
 \nn T^4 &=& \sin\theta_1\sin\theta_2\sin\theta_3 .
\end{eqnarray}
One finds that these functions satisfy (see also \cite{Sadri:2003mx})
\begin{equation}
[T^A,T^B,T^C] = \pi\varepsilon^{ABCD}T^D ,
\end{equation}
and also $\Tr(T^A,T^B) = \frac{1}{4}\delta^{AB}$. Thus the functions $T^A$ provide a representation of ${\cal A}_4$ inside ${\cal C}^{\infty}(S^3)$.

We now return to the fuzzy funnel solution. In the infinite $n$
limit, we expect that the fuzzy sphere loses its `fuzziness.'
We define the physical radius to be ({\it cf} (\ref{Rdef}))
\begin{equation}
R^2 = \frac{\Tr(X^A,X^A)}{T_2} .
\end{equation}
Following the calculations of the previous section, we find that the energy of the fuzzy funnel is
\begin{equation}
E = T_5 \int d^5x ,
\end{equation}
which exactly reproduces the tension of an M5-brane.

It seems natural to propose that this three-algebra is the large $n$
limit of the finite dimensional three-algebras that describe $n$
M2-branes. It is tempting to further speculate that the
three-manifold $\Sigma$ should somehow be identified with the
worldvolume of the M2-branes (or possibly with the worldvolume of an open
M2-brane that plays a role analogous to the one that open strings
play in the definition of D-branes). The gauge symmetries are then
simply the area preserving diffeomorphisms of the M2-brane
worldvolume. Note that the gauge field is non-dynamical with a
Chern-Simons-like kinetic term, and that this is consistent with
identifying it with the metric in three dimensions.
Is it intriguing to note that area-preserving diffeomorphism have
previously been associated with the gauge symmetry of M2-branes
\cite{de Wit:1988ig}.

Furthermore, we observe that if $\Sigma$ is the worldvolume of the
M2-branes, then under a parity transformation the triple product
(\ref{Nambu}) changes sign. This would then explain why one
needs to have $f^{abcd}\to -f^{abcd}$ in the finite dimensional
cases to preserve parity.

\section{\sl Conclusions}

In this paper we have analyzed various physical aspects of the
multiple M2-brane Lagrangian proposed in \cite{Bagger:2007jr}. In
particular, we discussed the quantization of $f^{abcd}$ that is
required by the quantum theory, the vacuum moduli space of the
simplest example, and various features of fuzzy sphere vacua and
fuzzy funnels. We also proposed a natural infinite $n$
three-algebra, and showed that it correctly produces the energy
density of a fuzzy-funnel solution, with no arbitrary parameters. In
so far as we have been able to check, the theory is consistent with
all expectations. It would also be interesting to compare this
model with predictions from the BFFS matrix model description of
M-theory \cite{Banks:1996vh}.

We believe that the most pressing open issue is obtaining an
infinite class of three-algebras that can represent an arbitrary
number of M2-branes. There is a large literature on related algebras
that arise from quantization of the Nambu bracket, starting with the
work of \cite{Nambu:1973qe,Takhtajan:1993vr}. However, much of this
literature imposes slightly different conditions on the
triple-product, such as a Leibnitz property that we do not require
or a generalized Jacobi identity that is weaker than the fundamental
identity (for example see \cite{SheikhJabbari:2004ik}). With such a
class of three-algebras, one would presumably be able to analyze the
vacuum moduli spaces and deduce the infamous relation $N = n^{3/2}$.
(For an alternative derivation, see \cite{BC}.)

Finally, we note that in this paper we have restricted our attention to
the algebraic structure presented in \cite{Bagger:2006sk,Bagger:2007jr}.  However, as shown in the appendix, there is an equivalent definition that was
introduced in \cite{Gustavsson:2007vu}.  The relation between the two is worth exploring in greater detail.

\section*{\sl Acknowledgements}

We would like to thank  A.\ Gustavsson for email correspondence and
M.\ van Raamsdonk for comments. JB is supported in part by the US
National Science Foundation, grant NSF-PHY-0401513. NL is supported
in part by the PPARC grant PP/C507145/1 and the EU grant
MRTN-CT-2004-512194.

\section*{\sl Appendix: Equivalence with Ref.~\cite{Gustavsson:2007vu}}

In \cite{Gustavsson:2007vu}, Gustavsson presented an algebraic
structure in which there are two vector spaces $\cal A$ and $\cal B$.
For $\alpha,\beta \in {\cal A}$ and $A,B\in {\cal B}$, he considered
bi-linear products of the form
\begin{eqnarray}\label{brakets}
\nn \langle\alpha ,\beta \rangle &=& -\langle\beta,\alpha \rangle \in {\cal B}  \\
  (A,\alpha) & \in& {\cal A} \\
\nn [A,B]&=&-[B,A]\in {\cal B} ,
\end{eqnarray}
as well as the `associative' condition
\begin{equation}\label{ass}
(\langle\alpha,\beta\rangle,\gamma) =
(\langle\beta,\gamma\rangle,\alpha).
\end{equation}
He then imposed the `Jacobi' identities
\begin{eqnarray}\label{Jacs}
\nn \langle (A,\alpha),\beta\rangle - \langle (A,\beta),\alpha\rangle  &=& [A,\langle \alpha,\beta\rangle ] \\
  (A,(B,\alpha)) -(B,(A,\alpha))&=&([A,B],\alpha) \\
\nn  [[A,B],C]+[B,[A,C]] &=& [A,[B,C]].
\end{eqnarray}
Note that the final condition
is simply the statement that $\cal B$ is a Lie algebra. With these
structures we can construct a tri-linear product on $\cal A$,
\begin{equation}
[\alpha,\beta,\gamma]\equiv (\langle\alpha ,\beta \rangle,\gamma).
\end{equation}
Note that this triple product is manifestly antisymmetric in
$\alpha\leftrightarrow\beta$ and the associative condition
(\ref{ass}) further implies that it totally antisymmetric in
$\alpha,\beta,\gamma$.

We will first show that Gustavsson's structure satisfies the fundamental identity cited in \cite{Bagger:2007jr}. Using the second Jacobi identity, we find
\begin{eqnarray}
\nn  [\alpha,\beta,[\gamma,\delta,\epsilon]] &=& (\langle\alpha ,\beta \rangle,(\langle \gamma ,\delta\rangle,\epsilon)) \\
&=&(\langle\gamma ,\delta \rangle,(\langle \alpha
,\beta\rangle,\epsilon)) +
([\langle\alpha ,\beta \rangle,\langle \gamma ,\delta\rangle],\epsilon)\\
\nn&=&[\gamma ,\delta,[\alpha ,\beta ,\epsilon]] +([\langle\alpha
,\beta \rangle,\langle \gamma ,\delta\rangle],\epsilon),
\end{eqnarray}
The second term on the right-hand side can be rewritten using the first Jacobi identity as
\begin{eqnarray}
\nn[\langle\alpha ,\beta \rangle,\langle \gamma ,\delta\rangle]
&=&\langle (\langle\alpha,\beta\rangle, \gamma),\delta \rangle -
\langle
(\langle\alpha,\beta\rangle, \delta),\gamma \rangle\\
&=&\langle [\alpha,\beta, \gamma],\delta \rangle - \langle
[\alpha,\beta, \delta],\gamma \rangle,
\end{eqnarray}
and hence
\begin{eqnarray}
([\langle\alpha ,\beta \rangle,\langle \gamma
,\delta\rangle],\epsilon) &=&[[\alpha,\beta,
\gamma],\delta,\epsilon]- [[\alpha,\beta , \delta],\gamma,\epsilon].
\end{eqnarray}
Thus we see that
\begin{equation}
[\alpha,\beta,[\gamma,\delta,\epsilon]] =[\gamma ,\delta,[\alpha
,\beta ,\epsilon]]+[[\alpha,\beta, \gamma],\delta,\epsilon] +
[\gamma,[\alpha,\beta, \delta],\epsilon],
\end{equation}
which is the fundamental identity.  This proves that the algebraic structure
introduced in \cite{Gustavsson:2007vu} satisfies the algebraic
condition in \cite{Bagger:2007jr}.

To show equivalence, we also need to prove the other way around.
Therefore we start with the algebraic structure used in
\cite{Bagger:2007jr}, consisting of a single vector space $\cal A$
with elements $\alpha,\beta,...$, and a tri-linear totally
antisymmetric product
\begin{equation}
[\alpha,\beta,\gamma] \in {\cal A}
\end{equation}
that satisfies the fundamental identity
\begin{equation}\label{FID}
[\alpha,\beta,[\gamma,\delta,\epsilon]] =
[[\alpha,\beta,\gamma],\delta,\epsilon]+[\gamma,[\alpha,\beta,\delta],\epsilon]]
+[\gamma,\delta,[\alpha,\beta,\epsilon]],
\end{equation}
and show that how to construct bilinear products that obey the relations (\ref{brakets}) -- (\ref{Jacs}).

For $X\in {\cal A}$, we can define a vector space of
linear maps from $\cal A$ to itself, generated by
\begin{equation}
as_{\alpha,\beta}(X)  = [\alpha,\beta,X],
\end{equation}
This is the space ${\cal B}$, with elements generated by $A = a s_{\alpha,\beta}$.  (In \cite{Bagger:2007jr}, this space was denoted by $\cal G$.)
One sees, using the fundamental identity, that
\begin{equation}\label{id}
[\alpha_1,\beta_1,[\alpha_2,\beta_2,X]] -(1\leftrightarrow 2)
 = [[\alpha_1,\beta_1,\alpha_2],\beta_2,X] +
 [\alpha_2,[\alpha_1,\beta_1,\beta_2],X].
\end{equation}
This shows that $as_{\alpha_1,\beta_1} \circ
as_{\alpha_2,\beta_2}-as_{\alpha_2,\beta_2}\circ
as_{\alpha_1,\beta_1}$ is again an element of $\cal B$.  In fact,
linear maps of a vector space to itself are associative under
composition, so $\cal B$ is a Lie algebra using the
ordinary commutator, satisfying the Jacobi
identity (which is the final condition in (\ref{Jacs})).
Equation (\ref{id}) also shows that
\begin{equation}
\label{gret} [{A_1},{A_2}](X)
=[[\alpha_1,\beta_1,\alpha_2],\beta_2,X] +
[\alpha_2,[\alpha_1,\beta_1,\beta_2],X],
\end{equation}
where ${A_i}$ denotes the map $A_i(X) = as_{\alpha_i,\beta_i}(X)$.
Thus the right hand side is actually anti-symmetric in $A_1
\leftrightarrow A_2$.

The rest of algebraic structure introduced in
\cite{Gustavsson:2007vu} can be constructed as follows.
It is natural to define
\begin{equation}
\langle\alpha,\beta \rangle  = as_{\alpha,\beta},
\end{equation}
and
\begin{equation}
(A,\alpha) = as_A(\alpha).
\end{equation}
The condition (\ref{ass}) then follows from the antisymmetry of
$[\alpha,\beta,\gamma]$.  To prove (\ref{Jacs}), we compute
\begin{eqnarray}
\nn\langle (A,\alpha),\beta\rangle(X) - \langle
(A,\beta),\alpha\rangle (X)  &=&
[as_A(\alpha),\beta,X] - [as_A(\beta),\alpha,X] \\
\nn &=& [as_A(\alpha),\beta,X] + [\alpha,as_A(\beta),X]\\
 &=& [as_A,as_{\alpha,\beta}](X)\\
\nn &=& [as_A,\langle\alpha,\beta\rangle](X)
\end{eqnarray}
for arbitrary $X$.  This reproduces the first condition
in (\ref{Jacs}).  We then compute
\begin{eqnarray}
\nn (A,(B,\alpha)) - (B,(A,\alpha))&=&as_A (as_B(\alpha)) -as_B( as_A(\alpha))\\
   &=&[as_A,as_B](\alpha)\\
\nn   &=& ([A,B],\alpha),
\end{eqnarray}
which is the second condition in (\ref{Jacs}). Thus a three-algebra
that satisfies the fundamental identity also provides an example of
the algebraic structure in \cite{Gustavsson:2007vu}.  This proves that
the two approaches are, in fact, equivalent.

\end{document}